# Multiscale modelling in nuclear ferritic steels: from nano-sized defects to embrittlement.

N. Castin[1,*], G. Bonny[1], M.J. Konstantinović [1], A. Bakaev[1], F. Bergner[2], C. Courilleau[4], C. Domain[3], B. Gómez-Ferrer[4], J.M. Hyde[5,6], L. Messina[7], G. Monnet[3], M.I. Pascuet[8], B. Radiguet[4], M. Serrano[9], L. Malerba[9]

[1] Studiecentrum voor Kernenergie – Centre D'Études de L'énergie Nucléaire (SCK CEN), NMS Unit, Boeretang 200, Mol, B2400, Belgium.
[2] Helmholtz-Zentrum Dresden-Rossendorf, Institute of Resource Ecology, Bautzner Landstr. 400, Dresden, 01328, Germany.
[3] EDF-R&D, Département Matériaux et Mécanique des Composants (MMC), Les Renardières, Moret sur Loing Cedex, F-77818, France.
[4] Groupe de Physique des Matériaux, Université et INSA de Rouen, UMR CNRS 6634, B.P. 12, Saint-Etienne Du Rouvray Cedex, 76801, France.
[5] National Nuclear Laboratory, Culham Science Centre, Abingdon, Oxfordshire, OX14 3DB, UK.
[6] Department of Materials, University of Oxford, Parks Road, Oxford, OX1 3PH, UK.
[7] CEA, DES, IRESNE, DEC, Cadarache F-13108 Saint-Paul-Lez-Durance, France.
[8] Gerencia Materiales - CAC, CNEA/CONICET, Godoy Cruz 2290, C1425FQB, CABA, Argentina.
[9] Centro de Investigaciones Energéticas, Medioambientales y Tecnológicas (CIEMAT), Avda. Complutense 40, Madrid, 28040, Spain.
[*] Corresponding author: ncastin@sckcen.be; nicolas.m.b.castin@gmail.com

Manuscript finalized on 25 April 2022

## Abstract

Radiation-induced embrittlement of nuclear steels is one of the main limiting factors for safe long-term operation of nuclear power plants. In support of accurate and safe reactor pressure vessel (RPV) lifetime assessments, we developed a physics-based model that predicts RPV steel hardening and subsequent embrittlement as a consequence of the formation of nano-sized clusters of minor alloying elements. This model is shown to provide reliable assessments of embrittlement for a very wide range of materials, with higher accuracy than industrial correlations. The core of our model is a multiscale modelling tool that predicts the kinetics of solute clustering, given the steel chemical composition and its irradiation conditions. It is based on the observation that the formation of solute clusters ensues from atomic transport driven by radiation-induced mechanisms, differently from classical nucleation-and-growth theories. We then show that the predicted information about solute clustering can be translated into a reliable estimate for radiation-induced embrittlement, via standard hardening laws based on the dispersed barrier model. We demonstrate the validity of our approach by applying it to hundreds of nuclear reactors vessels from all over the world.

# 1 Introduction

Nuclear power plant lifetime extension and new builds are assets to fight climate change [1]. The main lifetime limiting factor for nuclear power plants is the integrity of the steels composing the pressure vessel, as their mechanical properties degrade under neutron irradiation. Reactor pressure vessel (RPV) steels in "western" nuclear power plants, as well as their Russian counterpart, "VVER" steels, become increasingly brittle with increasing irradiation dose at operating temperature [2]. Ferritic steels are also foreseen in new generations of nuclear reactors, where they will be subjected to harsh environments never experienced at large industrial scale before, noticeably in terms of temperature and neutronic dose. Ferritic/martensitic (F/M) steels, candidate materials for fourth generation fission reactors and fusion applications [3, 4, 5], are also subjected to radiation-induced embrittlement. Systematic microstructural investigations on all these steels to determine the features of potentially embrittling defects are clearly not viable. By understanding the fundamental processes that take place in materials under irradiation [6], however, it is possible to develop models that provide this information and enable high quality predictions, which can be used in support of degradation monitoring procedures and safe plant lifetime extension. A model of this type is proposed in this work.

The most challenging aspect of the problem is that it all starts at the nanometer [7], or even at the atomic [8], scales, with consequences on the integrity of massive components that weigh several tons. It is indeed well-established that the degradation of most structural materials under irradiation unfolds from the formation of nano-sized defects (diameter $D$ = 1-4 nm) that are finely dispersed in the bulk (number density $N$ = $10^{22}$-$10^{24}$ m$^{-3}$) and cause hardening and ensuing embrittlement, at temperatures relevant to operating light water reactors (LWRs) [3, 9]. Such defects result from the diffusion and interaction of radiation-generated point defects between themselves, with the pre-existing microstructure and with alloying elements and impurities [8, 10]. In the specific case of ferritic steels, atomic transport mechanisms [11, 12] enrich point-defect clusters with Ni, Mn, Si, P and Cu [13]. Except in the case of the last element, enrichment is largely proportional to the abundance of each element in the matrix. This is revealed by atom probe tomography (APT) techniques [7, 13-17], as illustrated in Fig. 1, and by small angle neutron scattering (SANS) techniques [18]. These nano-sized clusters of solute elements obstruct the motion of dislocations, thereby hindering the blunting of crack tips. The dispersed-barrier hardening model [19] relates the increase $\Delta\sigma_y$ of the yield strength (hardening) with $\sqrt{ND}$ of the nano-clusters. In turn, the shift $\Delta T$ in the ductile-to-brittle transition temperature (DBTT) is correlated to $\Delta\sigma_y$ [20]. (The DBTT is the main parameter that is used at industrial level to define the lifetime of the vessel.) Thus, if $N$ and $D$ are known for a given steel, irradiated under known conditions, the corresponding $\Delta T$ can be estimated.

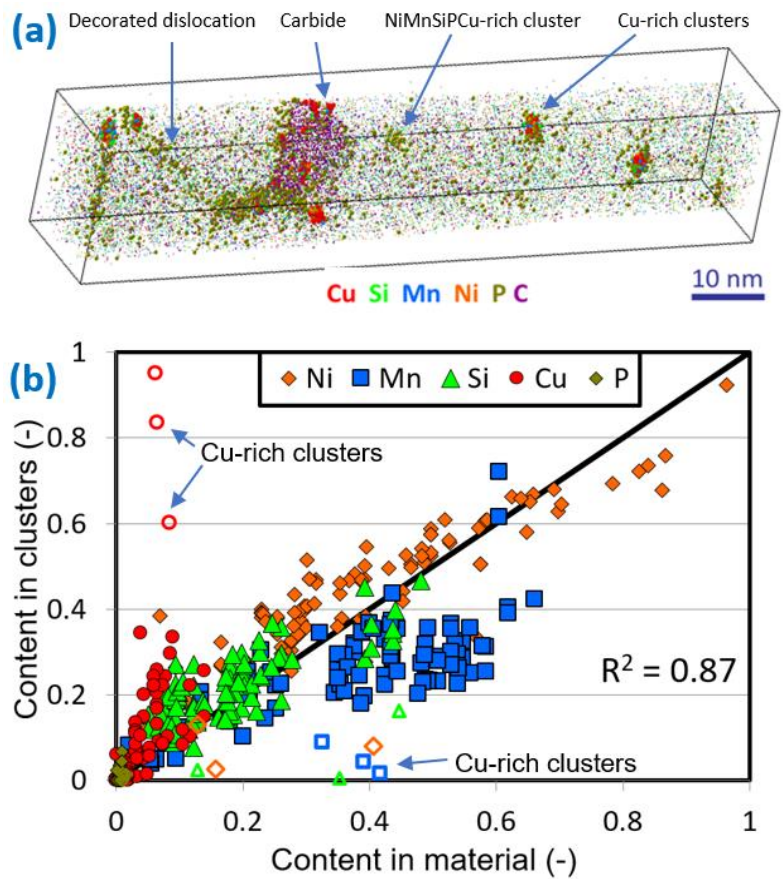

*Figure 1 – APT analysis of irradiated RPV steels. (a) 3-D atomic map reconstructed from a Cu-rich irradiated RPV steel. Several radiation-caused features are seen. Typical analyzed volumes contain 2-10 million atoms; (b) correlation between the chemical content in solute clusters in bulk and nominal concentration. Chemical contents are calculated in relative terms, excluding the Fe matrix atoms. Open symbols denote Cu-rich precipitates in 3 different materials [21-23], while closed symbols denote NiMnSiPCu-rich precipitates in 74 different materials [13,14, 17, 24-34]. Note that Cu is the only element that strongly deviates from the linear correlation with the element abundance in the alloy. Mn also slightly deviates because it is a carbide former, so part of the content of this element is found in carbides.*

Thermodynamic-driven precipitation of Cu and Cr occurs under irradiation below 300°C in materials containing more than ~0.1at.% Cu or ~9at.% Cr [5, 35]. In this class of materials, Cu-rich or Cr-rich precipitates are a clear cause of embrittlement [36] and thermodynamic models can be largely used to predict their formation. We address here the formation of clusters rich in Ni, Mn, Si and P in materials where no element exceeds the solubility limit of the binary alloy. The mechanism of solute cluster formation in this case is still debated. Is it an irradiation-enhanced process, dictated by thermodynamics [37, 38], or an irradiation-induced process that would not occur without irradiation [39, 40]? Recently, Ke *et al.* [38] developed a model based on the assumption that irradiation mainly accelerates the thermodynamics-driven kinetics of phase separation, due to the increased vacancy concentration. This model also included a corrective factor to account for possible "radiation-induced heterogeneous nucleation" of precipitates on point defect clusters produced in atomic displacement cascades. The application of the model revealed that the latter term is dominant by several orders of magnitude over the thermodynamic component, in some cases. This suggests that thermodynamics-driven processes are not dominant under irradiation.

APT provides several notable clues for the dominant radiation-induced driving force for solute clustering (Fig. 1):

a) Segregation is ubiquitous: in addition to finely dispersed clusters in the bulk, high concentrations of solutes (especially P [41]) occur along dislocation segments, at carbides and at grain boundaries.
b) Two classes of solute clusters can be distinguished by their chemical compositions. The composition of many solute clusters strongly correlates with the overall composition of the material (closed symbols in Fig. 1b). In contrast, in materials with high enough Cu or Cr content, Cu-rich [21, 35] (open symbols in Fig. 1b) or Cr-rich [5] clusters are found.
c) Clusters are dilute, as they contain a significant fraction of Fe [14, 16, 17]. APT often finds iron contents ≥50% [27], while other techniques (electron microscopy, SANS) [18, 27] suggest instead a content ≈20% Fe.

These clues advocate for a continuous process of solute deposition on a large density of nucleation sites, triggered and sustained by the continuous fluxes of vacancies and self-interstitials due to irradiation. This mechanism is different from classical precipitation by nucleation and growth. It is the theoretical foundation that we rely on to present in this work a model describing the formation of NiMnSiPCu-rich clusters under irradiation.

The nuclear industry typically estimates the evolution of $\Delta T$ with the received neutron dose, for given steel and irradiation conditions, employing empirical regressions [42] that are fitted to experimental data. Even though these correlations are inspired by physics (e.g., explicit terms stand for the so-called "matrix damage", others for the effects of solute-rich clusters, etc.), their extrapolation capabilities depend entirely on the choice of the mathematical expression that is used. In this work, we aim at setting the bases for the definition of models that are more strongly rooted in physics. For this purpose, we develop a multiscale tool that describes the evolution of the microstructure under irradiation. We combine first principles calculations, molecular dynamics simulations and atomistic modelling within a single coarse-grained tool which links the atomic to the nanometer scales, allowing the simulated time to be extended up to decades of neutron irradiation. The tool is therefore driven by elementary, atomic-level reactions, which implement the above-discussed evidence for radiation-induced solute clustering. This tool is a generalization of a previous model [13] which, given material composition and irradiation conditions, predicts the evolution of $N$ and $D$, and thus of the related cluster volume fraction, $f$, with the received irradiation dose. We then employ classical hardening laws, also rooted in physics, to correlate this information with the corresponding $\Delta\sigma$ and the ensuing $\Delta T$. This step can be regarded as a shortcut in the classical multiscale approach [6] towards the desired macroscopic level, as we replace mesoscale methods such as dislocation dynamics, continuous field models or finite elements methods with simpler laws and the correlation between hardening and embrittlement. However, this route remains open for future work based on the knowledge of $N$ and $D$.

## 2 Materials and Methods

### 2.1 Materials and experimental databases

This work is focused on the following families of materials:
- **Base materials RPV steels:** The pressure vessels of LWR, i.e. pressurized (P) and boiling (B) water reactors (P/BWR) are made of low carbon (<0.9-1.3 at%), low-alloyed (> 95 at% Fe) bainitic steels. The main alloying elements common to all RPV steels are Mo (~0.3 at%), Ni (~0.5-1.0 at%), Mn (~0.5-1.5 at%), and Si (~0.2-0.8 at%), chiefly added for strengthening purposes. We consider here experimental data from samples irradiated at different dose rates: low and medium, corresponding to surveillance samples (~$2.5 \cdot 10^{-10}$ dpa/s and ~$3 \cdot 10^{-9}$ dpa/s), and high, irradiated in a material test reactor, MTR (~$10^{-7}$ dpa/s).
- **Weld materials RPV steels**: these materials are also bainitic steels, but their composition is different, typically they are characterized by higher Ni and Mn content and more impurities; in particular, this class includes steels with very high – for RPV – Ni and Mn content, i.e. up to 2 at%. The irradiation conditions considered here are similar to those of the RPV-BM.
- **Materials used for the RPV of Russian design nuclear reactors (VVER):** these are bainitic steels that contain, in addition to the alloying elements added to western Europe RPV steels, also Cr (~2.0-2.5 at%) and V (0.1-0.35 at%), to provide higher thermal ageing resistance through formation of several types of Cr- and V-containing carbides and carbonitrides. The Ni content varies in a wider range for these materials, as it can be either relatively high (~1.5 at%) or very low (< 0.1 at%). The samples considered here were irradiated in MTR at high dose rate (~$10^{-7}$ dpa/s) and in nuclear power plants at low dose rates ($10^{-10}$ to $10^{-9}$ dpa/s).

Two databases of experimental data were collected from the literature for the calibration and validation of our model:
- **The microstructure database** collects 104 experimental measurements with APT [13, 14, 16, 17, 21, 25-31, 32, 33, 39, 40, 43-49] from 55 irradiated materials (46 RPV steels, 6 VVER steels and 3 model alloys for RPV steels). For each case, there are 9 input data: the materials chemical composition (Ni, Mn, Si, P and Cu contents in at%), the irradiation conditions (temperature in °C and the dose rate in dpa/s), and the received neutron dose (in dpa, which is equivalent to the irradiation time given the dose rate). The output data are $N$ and $D$, and therefore $f$, of the solute clusters found in bulk, with the associated error bars, measured with APT when available.
- **The embrittlement database** collects 1707 experimental measurements from 359 irradiated materials (2 model alloys, 347 RPV steels, 10 VVER steels). Most of the data is taken from the PLOTTER database of international nuclear power surveillance [50], which is complemented with data found in the literature: see the microstructure database and the references provided above, further complemented with VVER data [51] and recent embrittlement data derived for high Ni model RPV steels [52]. Overall, we only retained the materials that correspond to a Cu content < 0.1 at%. The input data are identical as in the microstructure database, while the output data is the embrittlement $\Delta T$ in °C, and/or the hardening $\Delta\sigma_Y$ in MPa. The specific range of uncertainty associated with each of these measurements should ideally be provided case by case, from the experimental procedure that is applied. Since precise information is scarce, we assume that the uncertainty on $\Delta T$ is ± 25°C.

Both databases are provided as supplementary materials. In both cases, original neutron fluence and neutron flux data provided in n cm$^{-2}$ or in n cm$^{-2}$ s$^{-1}$, respectively, are converted into displacement-per-atom, dpa, or dpa s$^{-1}$ using a constant conversion factor: 1 dpa = 6.67 10$^{20}$ n cm$^{-2}$ [53]. In addition, this work aims at predicting the embrittlement $\Delta T$ induced by radiation, but experimental data found in the literature often concern the increase in the yield strength (hardening) $\Delta\sigma_y$ instead. As both are correlated [20], hardening data is here converted into an estimated embrittlement value with the relation:

$$\Delta T (°C) = 0.5913\ \Delta\sigma_y\ (MPa) \tag{1}$$

The coefficient of proportionality was derived from the experimental data in the microstructure database. It is clear that the use of a constant conversion factor both from fluence to dpa and from hardening to embrittlement will be a source of scatter; however, as will be seen, the overall uncertainty remains acceptable.

## 2.2 Multiscale model for microstructure evolution under irradiation

In this section, we describe our physical model aimed at predicting $N$, $D$ and $f$ of the solute clusters, as functions of the material chemical composition (Ni, Mn, Si, P and Cu contents in at%), the irradiation conditions (temperature and dose rate), and the received neutron dose (equivalent to the irradiation time, given the dose rate).

The novelty of our approach compared to standard theories (such as, e.g., in [38]) is to explicitly account for the radiation-induced mechanisms discussed above, relying on elementary processes of diffusion, transport and binding at the atomic level. It is implemented through the coarse-grained object kinetic Monte Carlo (OKMC) computational method: see the schematic representation in Fig. 2. The fundaments of the model were defined in [13], where we demonstrated that the dominant mechanism for nano-sized solute cluster formation is the interaction of solute atoms with self-interstitial defects generated under irradiation. The OKMC simulation volume (in this work 270 $a_0$ × 300 $a_0$ × 420 $a_0$, thus including 68 million atoms in a perfect bcc lattice) is initially filled with the number of Ni, Mn, Si, P and Cu atoms that correspond to the targeted chemical composition. It is noteworthy that Cr atoms, that are present in the VVER steels in significant proportion, are not explicitly included in the simulation box, but their effect on the diffusivity of SIA defects is taken into account via an established grey alloy approximation [54, 55]. Next, primary point-defects (vacancies, self-interstitials and their clusters) are injected at the right frequency (according to the specified neutron dose rate) using libraries of atomic collision cascade debris previously derived from molecular dynamics. Key features of the model are solute transport phenomena induced by the injected point-defects, and their deposition at sinks. When single vacancies or self-interstitials encounter solute atoms dissolved in the matrix, they form stable pairs, with given binding energies. These pairs migrate with given activation energies towards sinks.

The characteristic energy values were calculated using density functional theory (DFT) methods [11, 12]. In addition to pre-existing microstructural features, such as dislocations and grain boundaries, point-defect clusters also act as sinks, especially self-interstitial atom clusters (denoted as C1 in Fig. 2). These are immobilized because of strong binding interactions with solute atoms accumulated round them, thereby acting as nuclei for further solute clustering.

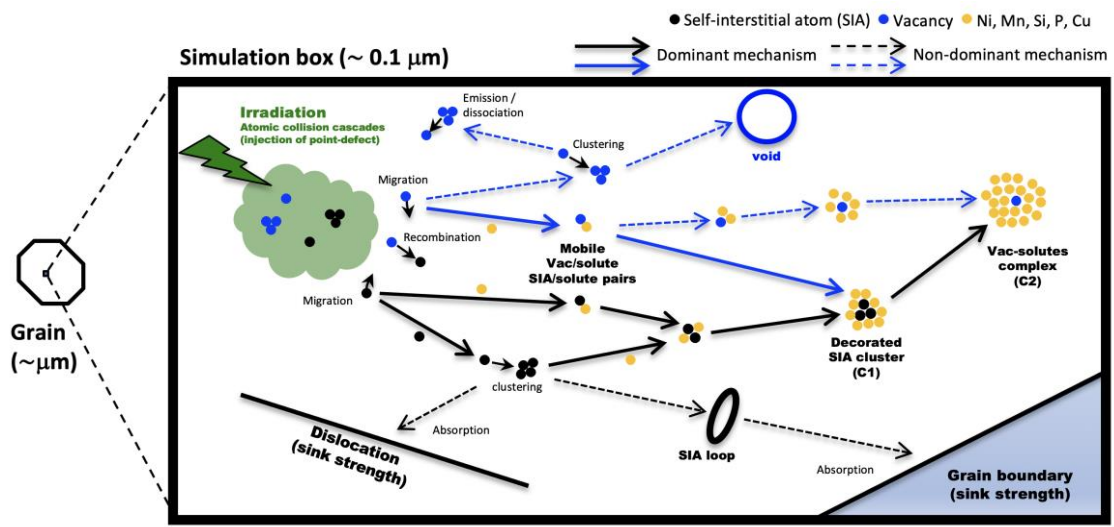

*Figure 2 – Schematic description of the object kinetic Monte Carlo method. From left to right, after the production of damage in collision cascades, elementary events take place, which are described as thermally-activated processes of diffusion, dissociation and transport by point-defects, all parameterized here based on first principle calculations. They spontaneously lead to the formation of solute rich clusters: complexes C1 and C2 contain a sufficient number of atoms to be resolvable with APT. The dominant mechanism [13] of formation is indicated by solid line arrows, while dashed arrows indicate secondary mechanisms which are, formally, not excluded from the model.*

While C1 self-interstitial complexes can be safely considered as indissoluble, the dissolution of C2 vacancy-solute complexes (Fig. 2) needs an appropriate description. The corresponding reactions are depicted in Fig. 3. Dissolution is a thermally activated process that takes place via two possible paths: emission of a single vacancy (henceforth denoted as "Va"), or emission of a vacancy-solute pair (henceforth "VaSo", where So is either Ni, Mn, Si, P or Cu). We estimate the dissociation energy $E_{diss}^{(Va,VaSo)}$ with the well-established approximate relation:

$$E_{diss}^{(Va,VaSo)} = E_m^{(Va,VaSo)} + E_b^{(Va,VaSo)} \qquad (2)$$

Here, $E_m^{(Va,VaSo)}$ is the migration energy of a Va or VaSo pair in bulk Fe, which is known from DFT calculations [11]. In contrast, the binding energy $E_b^{(Va,VaSo)}$ of a Va or a VaSo pair to C2 complexes requires special attention. In the absence of a theoretical framework, the approach of Ref. [13] can be used: they are determined by individual fitting, taking, for each given material, the binding energy that allows the OKMC model predictions for *N*, *D* and *f* to best match the experimental values from APT. However, in this way the model can only be applied to materials for which detailed microstructural data are available, such as those in the microstructure database described in section 2.1. It cannot be applied to the nuclear power plant surveillance materials collected in the PLOTTER database [50] because, to our knowledge, no APT analysis is available for these materials. The goal of the present work is to make such microstructural analysis unnecessary for the application of the model, given that in most cases it is not given: the model should be applicable with input variables that are always given, i.e. material composition and irradiation conditions. Therefore, as described in section 2.3, in this work we remove the limits of applicability of the OKMC method and propose, based on atomistic modelling and first principle calculations, a general recipe to estimate $E_b^{Va}$ and $E_b^{VaSo}$ *a priori*, as functions of the chemical composition and the size of the solute clusters, as schematically shown in Fig. 3 and described next.

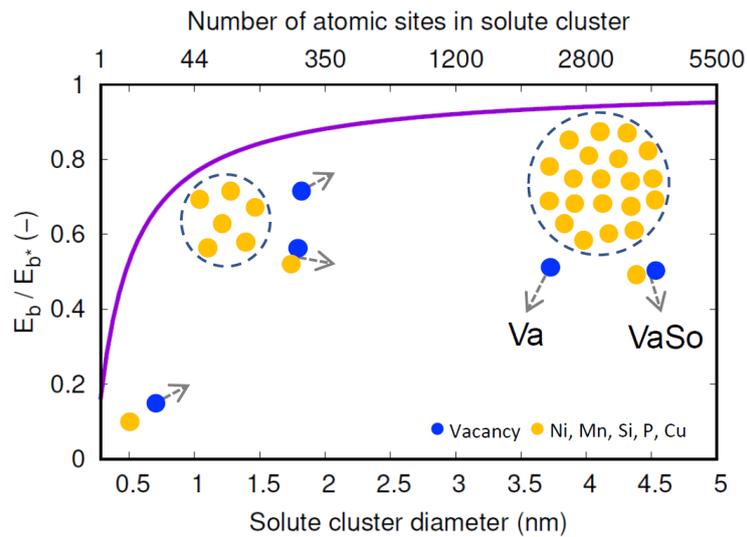

Figure 3 – Schematic description of how the binding energy $E_b$ of vacancies (Va) and vacancy-solute pairs (VaSo) to complexes C2 depend on cluster size. Values are normalized to the saturated value $E_{b^*}$.

### 2.3 Determination of key binding energies from atomistic modelling.

Small vacancy-solute complexes located on the left-hand side of Fig. 3 have been studied using DFT [13]. On the right-hand side of the figure, the other limiting case is a large complex, where the effects of the interface with the Fe matrix do not evolve with size any longer. The corresponding "saturated" binding energy values are denoted as $E_{b^*}^{Va}$ and $E_{b^*}^{VaSo}$. Intermediate cases can be interpolated using the $V_{C2}^{-1/3}$ capillary law, where $V_{C2}$ is the volume of the complex C2. The OKMC model is therefore fully parameterized once $E_{b^*}^{Va}$ and $E_{b^*}^{VaSo}$ are known, which solely depend on the chemical composition of the solute clusters.

To calculate $E_{b^*}^{Va}$ and $E_{b^*}^{VaSo}$, we need a Hamiltonian for an atomic system that contains the Fe, Ni, Mn, Si, Cu and P species. This is offered by the DFT-rooted cluster expansion (CE) model derived in [56]. Given a chemical composition of the C2 complex, we first evaluated $E_{b^*}^{Va}$ and $E_{b^*}^{VaSo}$ using the method developed in [57], which enables the calculation of defect energetics in random alloys on a rigid lattice based on an exact mean field approach. This corresponds to assuming that the atoms are randomly distributed inside the complexes, which is a fair assumption, given their mechanism of formation by segregation on point-defect clusters rather than via precipitation as phases. We applied this method for the materials listed in the microstructure database. Regarding the evidence shown in Fig. 1b, we assume that the relative chemical contents in Ni, Mn, Si, P and Cu are identical to those derived from the overall chemical content of the material. Finally, the Fe content in the solute clusters can be varied between 20 at% and 50 at% to cover the limiting cases between experimental evidence from SANS and APT, respectively.

## 2.4 Determination of embrittlement from solute clustering

According to the dispersed barrier hardening model [19], the radiation hardening $\Delta\sigma_y$ is a function of the density and the size of the array of obstacles to the motion of dislocations. In this work, we consider that the obstacles are the nano-sized solute clusters and apply the model originally proposed by Bacon, Kocks and Scattergood (BKS) [58], as formulated in [59], which considers the elastic interactions between segments of bowed-out dislocations by a population of randomly distributed obstacles. Accounting for Eq (1), we write:

$$\Delta T^{(BKS)} = 0.5913\, M\, \alpha_{BKS}^{3/2} \left(\frac{ln(\overline{D}/b)}{ln(l/b)}\right)^{3/2} \frac{ln(l/b)}{2\pi}\, \mu b \sqrt{ND} \qquad (3)$$

Here, M is Taylor's factor (taken to be 3), $l$ is defined as the free distance between solute clusters ($l = 1/\sqrt{ND} - D$), $\overline{D}$ is a harmonic sum ($\overline{D} = lD/(l+D)$), $\mu$ is the shear modulus and $b$ is the dislocation's Burgers vector module. The factor $\alpha_{BKS}$ accounts for the strength of solute clusters impeding the flow of dislocations under the applied strain. Following Friedel's interpretation [59], $\alpha_{BKS}^{3/2}$ can be directly related to the cosine of the bowing angle of the dislocations.

The solute clusters diameter $D$ is typically observed to vary within narrow ranges of values in databases of irradiated ferritic steels, at least in the range of temperatures and neutron doses that were studied the most. Considering that the volume fraction $f \approx ND^3$ and therefore $\sqrt{ND} \approx \sqrt{f}/D$, $D$ being almost constant, for simplicity and seeking for more accurate regressions, some authors [14, 15, 60] preferred to rely on this measurement of solute clustering directly. We hence write:

$$\Delta T^{(VF)} = \alpha_{VF} \sqrt{f} \qquad (4)$$

Here, the superscript (VF) stands for "volume-fraction-based model". Given our embrittlement database described in section 2.1 which provides values for $\Delta T$, and having $\sqrt{ND}$ and $\sqrt{f}$ either provided by our microstructure database or estimated by our theoretical model described in section 2.3, numerical values for $\alpha_{BKS}$ and $\alpha_{VF}$ can be estimated from linear regressions. We should emphasize that, for the sake of practicality, we adopt here a specific convention to define the volume fraction $f$, due to the fact that the latter is not systematically reported in APT examination work. We thus define $f = \frac{\pi}{6} ND^3$, irrespective of whether $N$ and $D$ are calculated using the OKMC model, or obtained from APT measurement.

# 3 Results

## 3.1 Calibration and validation of the microstructure evolution model

Following the methodology described in section 2.3, we performed atomistic calculations of $E_{b*}^{Va}$ and $E_{b*}^{VaSo}$ for the materials listed in the microstructure database described in section 2.1. We found that the Fe content can be varied between 20 at% and 50 at% without qualitatively affecting the results. We also found that $E_{b*}^{Va}$ and $E_{b*}^{VaSo}$ are strongly correlated with the relative Ni content in the materials, while the content in the other chemical elements play only a secondary role. Looked from that perspective, the values fall within a narrow range of variability. This is a crucial result of the physical atomistic model, as it greatly simplifies the form of the correlation we are looking for. As shown in Fig. 4, it is accordingly possible to write the following estimate functions, which hold for all chemical species (S = Ni, Mn, Si, P or Cu):

$$E_{b*}^{Va}\big|_{CE} \cong 1.2 - 0.4\, x_{Ni} \pm 0.01 \; \text{(eV)} \tag{5}$$

$$E_{b*}^{VaSo}\big|_{CE} \cong 1.56 - 0.85\, x_{Ni} \pm 0.15\, (1 - x_{Ni}) \; \text{(eV)} \tag{6}$$

Here, $x_{Ni}$ is the relative Ni content defined as $C_{Ni} / (C_{Ni} + C_{Mn} + C_{Si} + C_P + C_{Cu})$ and $C_{Ni},\ldots,C_{Cu}$ are the chemical contents in the materials. The difference between the $E_{b*}^{Va}\big|_{CE}$ curve and the $E_{b*}^{VaSo}\big|_{CE}$ curve for a given Ni content is consistent with estimates deduced from PAS experiments [61]. The estimates in Eq. (5-6) could in principle be used directly to parameterize OKMC simulations. However, the superposition of independent migration and binding energies in Eq. (2) does not account for dynamic effects that may affect the true value of $E_{diss}^{(Va,VaSo)}$. To improve accuracy without undue complications, corrections are introduced only on $E_{b*}^{VaSo}$. In the OKMC model, we therefore assume that:

$$E_{b*}^{Va}\big|_{OKMC} = E_{b*}^{Va}\big|_{CE} \tag{7}$$

In contrast, a better estimate for $E_{b*}^{VaSo}$ is derived from another function of the material chemical composition, determined as follows. OKMC simulations were performed for the materials listed in the microstructure database, in the corresponding irradiation conditions. In each case, $E_{b*}^{Va}$ was taken from Eq. (7) and $E_{b*}^{VaSo}$ was varied between $E_{b*}^{Va}\big|_{CE}$ and $E_{b*}^{VaSo}\big|_{CE}$, used as reference physical values, to determine the one that leads to the closest predictions of the solute clusters' volume fraction $f$, as compared to the APT data. The so-fitted values for individual materials are shown by dots in Fig. 4. The error bars indicate the typical range of tolerance for $E_{b*}^{VaSo}$ to achieve a good match with APT. Clearly, the optimal $E_{b*}^{VaSo}$ values can be approximated by a regression of $x_{Ni}$, similarly to the cluster expansion finding in Eq. (5) and Eq. (6). We therefore derived an overall single regression that relates $E_{b*}^{VaSo}$ with $x_{Ni}$, which minimizes the average prediction error by the OKMC model for $\sqrt{f}$ within the whole microstructure database, as follows:

$$E_{b*}^{VS}\big|_{OKMC} \cong -5.30 x_{Ni}^2 + 0.109\, x_{Ni} + 1.178 + E_{b*}^{VS}\big|_{Cu} \pm 0.025 \ (\text{eV}) \qquad (8)$$

Here, $E_{b*}^{VS}\big|_{Cu}$ is an additional binding energy term which accounts for the presence of Cu. We fitted a linear ramp step function that is effective from a moderate Cu content $C_{Cu}$:

$$E_{b*}^{VS}\big|_{Cu} = 0 \qquad \text{if } (C_{Cu} < 0.05 \text{ at\%}) \qquad (9)$$
$$= (C_{Cu} - 0.05) * 0.035 \ (\text{eV}) \qquad \text{if } (0.05 \text{ at\%} \leq C_{Cu} \leq 0.07 \text{ at\%}) \qquad (10)$$
$$= 0.035 \ (\text{eV}) \qquad \text{if } (C_{Cu} > 0.07 \text{ at\%}) \qquad (11)$$

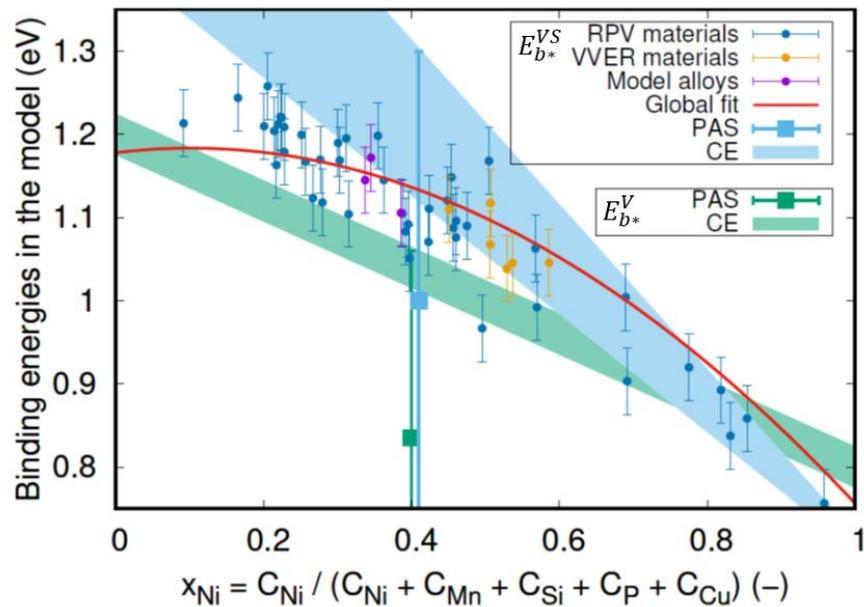

*Figure 4 – Saturated binding energy values estimated in different ways, as functions of the relative Ni content in the material. The acronyms PAS stands for positron annihilation spectroscopy (experimental measurement taken from [60]), and CE for cluster expansion (theoretical model based on density functional theory).*

With Eqs (7-11), the OKMC model is fully parameterized, and can therefore be applied to assess solute clustering in any RPV material (i.e. of given composition in terms of minor solutes), under any irradiation condition. Fig. 5 shows a comparison between the predictions achieved with the model and APT data for the materials listed in the microstructure database. In Fig. 5a, we see that a reasonable correlation exists between the predicted and experimental volume fraction $f$ of solute clusters, despite large scatter with the largest values. In Fig. 5b and Fig. 5c, we see that the OKMC model achieves this reasonable prediction of $f$ by slightly overestimating the experimental $N$ and, in general, underestimating the average $D$. The latter is predicted to vary within a much narrower range than observed experimentally. We interpret this discrepancy as the result of an expectable divergence in resolution between, on the one hand, an experimental technique where the smallest defects are invisible (APT), and, on the other, a theoretical model where all defects are accounted for.

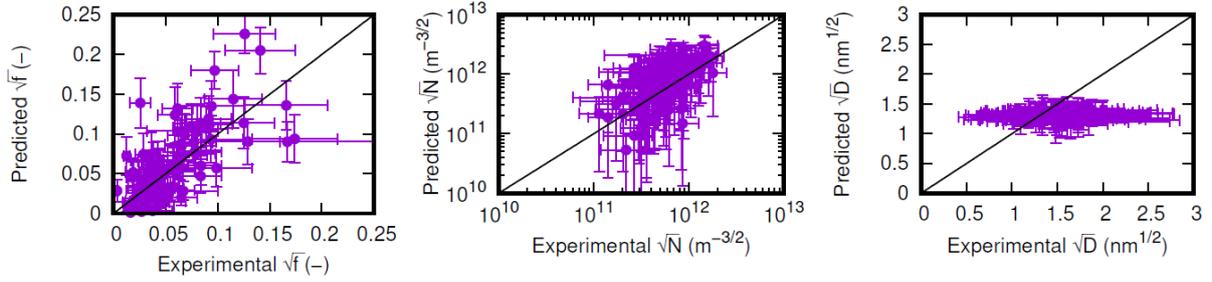

*Figure 5 – Comparison between the OKMC predictions and APT experimental data for the materials and irradiation conditions listed in the microstructure database.*

### 3.3 Predictions of radiation-induced embrittlement

Via the relations in Eq (2) and Eq (3), the coefficients $\alpha_{BKS}$ and $\alpha_{VF}$ can be determined from linear regressions using a given set of data about solute clusters ($\sqrt{ND}$), and the corresponding $\Delta T$.

We start by performing this exercise using the experimental data from APT collected in the microstructure database. The obtained coefficients are listed in Tab. 1, and the achieved predictions of $\Delta T$ are compared with the desired values in the top panels in Fig. 6. Both embrittlement laws appear to correlate the embrittlement $\Delta T$ with the amount of solute clustering on the average.

*Table 1 – Coefficients derived for embrittlement laws using the BKS model or the volume fraction based model. The coefficients are either derived from APT experimental data, or from OKMC estimates, for N and D of the solute clusters. In each case, the 95% interval of confidence (2$\sigma$) and the correlation coefficients $R_2$ are reported.*

|  | BKS model Eq. (3) | | | Volume fraction model Eq. (4) | | |
|---|---|---|---|---|---|---|
|  | $\alpha_{BKS}$ (-) | 2$\sigma$ (°C) | $R_2$ | $\alpha_{vf}$ (°C) | 2$\sigma$ (°C) | $R_2$ |
| **N and D From APT** | 0.478 ± 0.0923 | 41.9 | 0.80 | 1383 ± 129.8 | 42 | 0.80 |
| **N and D From OKMC** | 0.466 ± 0.0277 | 27.9 | 0.84 | 1550 ± 22.7 | 28.3 | 0.91 |

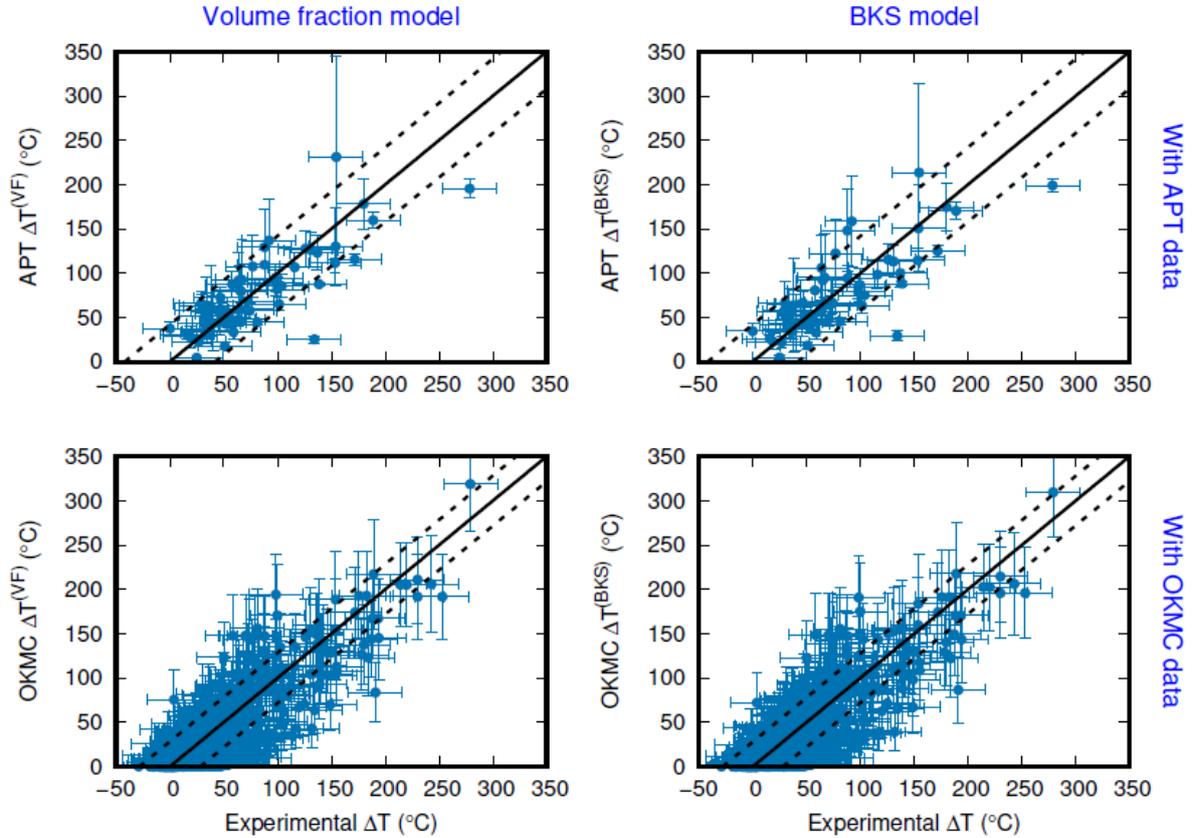

*Figure 6– Application of the embrittlement laws from Eq (2) (right panels) and Eq (3) (left panels) using the microstructure database (top panels) and the embrittlement database (bottom panels) described in section 2.1. Black lines are the ideal match, and the dashed lines show the 95% interval of confidence ($2\sigma$). The horizontal error bars on the dots is set to 25°C, while the vertical error bars are derived from the experimental uncertainty of the APT measurements or of the OKMC predictions.*

Next, we use the fitted $\alpha_{BKS}$ and $\alpha_{VF}$ coefficients of Tab. 1, to correlate the OKMC model predictions of *N*, *D* and *f* with the embrittlement. This can be now done without further calibration to all materials and their respective irradiation conditions collected in the embrittlement database. The comparison between OKMC predicted and target values is shown in the bottom panels of Fig. 6. Remarkably, both embrittlement laws in Eq (3) and Eq (4) work well for nearly all *ΔT* data-points, with high accuracy.

## 4 Discussion

Our results show that the linear relations between $\sqrt{ND}$ or $\sqrt{f}$ and *ΔT* as defined in Eq (3) and Eq (4) are general and can be applied to any RPV material via our physical model, without the need of APT data from time-consuming and costly microstructural examination. Importantly, these relations work well also for the high dose data points of the database, suggesting that the model enables accurate extrapolation to higher irradiation doses than those of surveillance. These simple embrittlement laws, coupled with $\sqrt{ND}$ or $\sqrt{f}$ computed using the OKMC model, can therefore be used to evaluate the integrity of the RPV in any of the world's operating nuclear power plants. Examples are shown in Fig. 7. It is worth noting that the rightmost irradiation dose of 0.4 dpa correspond to about 80 years of operation for pressurized-water reactors.

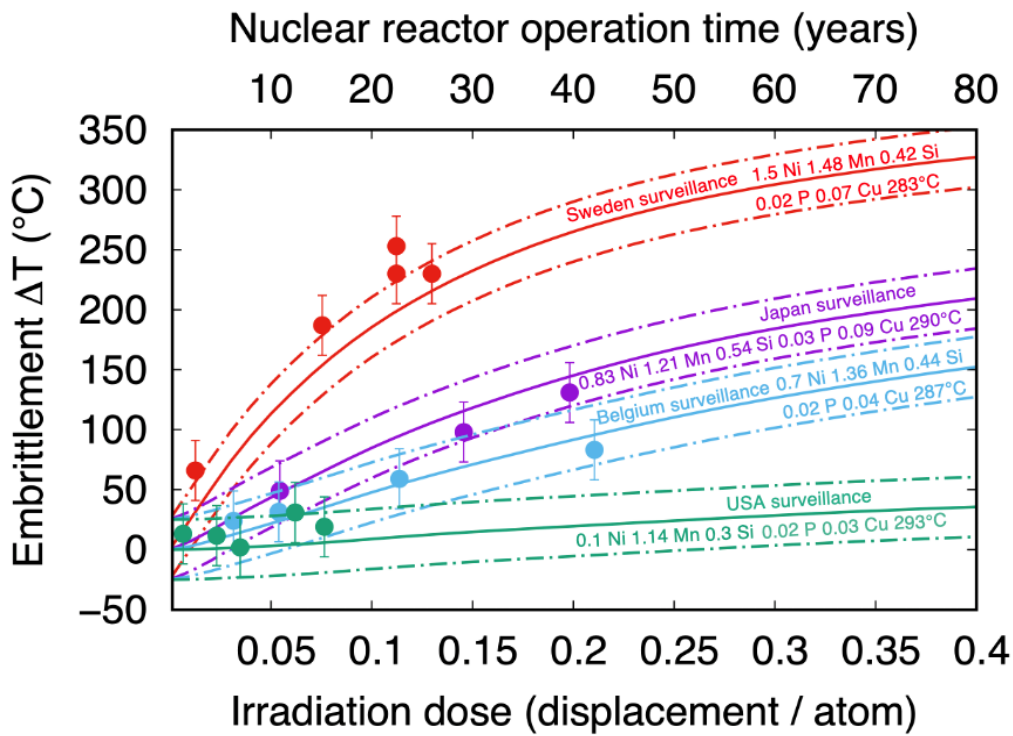

*Figure 7 – Physics-based law (volume-fraction model in Eq (3)) for radiation embrittlement applied to individual steels. Evolution with the received dose of the predicted (solid lines) and measured (dots) embrittlement ΔT for four individual steels (different colors). The 95% interval of confidence (±2σ = 27.5°C) is shown as well (dashed lines). Chemical compositions are given in at%. The irradiation dose of 0.4 displacement per atoms corresponds to about 80 years of operation for pressurized water reactors.*

To verify the robustness of our model, in Fig. 8 and Fig. 9, three industrial embrittlement-dose correlations are applied to the same embrittlement database as in Fig. 6 and compared with the current model's results. The generated predictions are fully listed in the supplementary materials (Part 2 in the "Embrittlement database" tab). The study included: (a) The EONY correlation [62], although thought for industrial use, is conceived as a physics-inspired model, by including embrittlement contributions from the so-called "stable matrix damage" and from Cu-rich precipitates; (b) The Regulatory Guide 1.99 Rev. 2 [63], in contrast, is a fully industrial trend curve with embrittlement contributions from a chemistry factor and a fluence function, which is currently the reference correlation in the US and not only for integrity assessment; (c) The ASTM E900-15 correlation [64], which can be regarded as another trend curve fitted to more recent surveillance data, including some for which the Regulatory Guide 1.99 Rev. 2 noticeably failed. These three embrittlement-dose correlations were numerically fitted to surveillance data. It is clear from Fig. 8 and Fig. 9 that all correlations are accurate in most cases. However, in some cases, supposedly outside of the range of fitting, the industrial trend curves clearly fail and the error of prediction is large (up to 300°C). In Fig. 9 the evolution of the prediction error is also plotted as a function of the three most influencing input variables of the models, i.e., the Cu content, the Ni content, the temperature, and the neutron fluence. We observe that:

- The industrial correlations fail to make accurate predictions when the Ni content is high (above ~1.5%). This is of course expected for very high Ni contents (> 3 w%) that are atypical compared to industrially-relevant steels: these materials have been chemically tailored for specific scientific purposes [14]. Ni contents between 1.5 wt% and 2 wt%, however, are industrially relevant. The model proposed in this work achieves overall more accurate predictions compared to the industrial trend curves for these materials.
- The industrial correlations commit the largest error of predictions when the neutron fluence is larger than $3 \cdot 10^{19}$ n cm$^2$ (equivalent to 0.05 dpa). On the contrary, the prediction error committed by the model proposed in this work is contained in a noticeably narrower range.
- The model proposed in this work performs with equal accuracy in the whole range of irradiation temperatures (from 255°C to 300°C), even though a slight bias towards underestimation is detectable in the lowest temperature range. More low temperature (<275ºC) experimental data would be required to perform a detailed study, but the model proposed in this works seems to be at least as accurate as the Regulatory Guide trend curve for temperatures < 275°C, while the EONY model is perhaps slightly more accurate in this range.

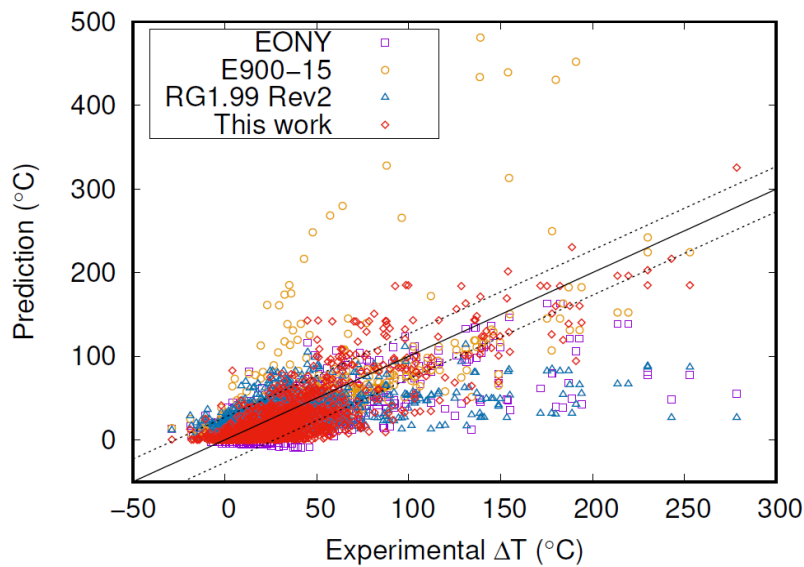

*Figure 8 – Accuracy of prediction of ΔT achieved by the EONY embrittlement correlation method [62], The Regulatory Guide 1.99 Rev. 2 [63], and the ASTM E900-15 correlation [64], compared with the predictions achieved in this work. The black line is the ideal match, and the dashed lines show the 95% interval of confidence (2σ = 37 °C ).*

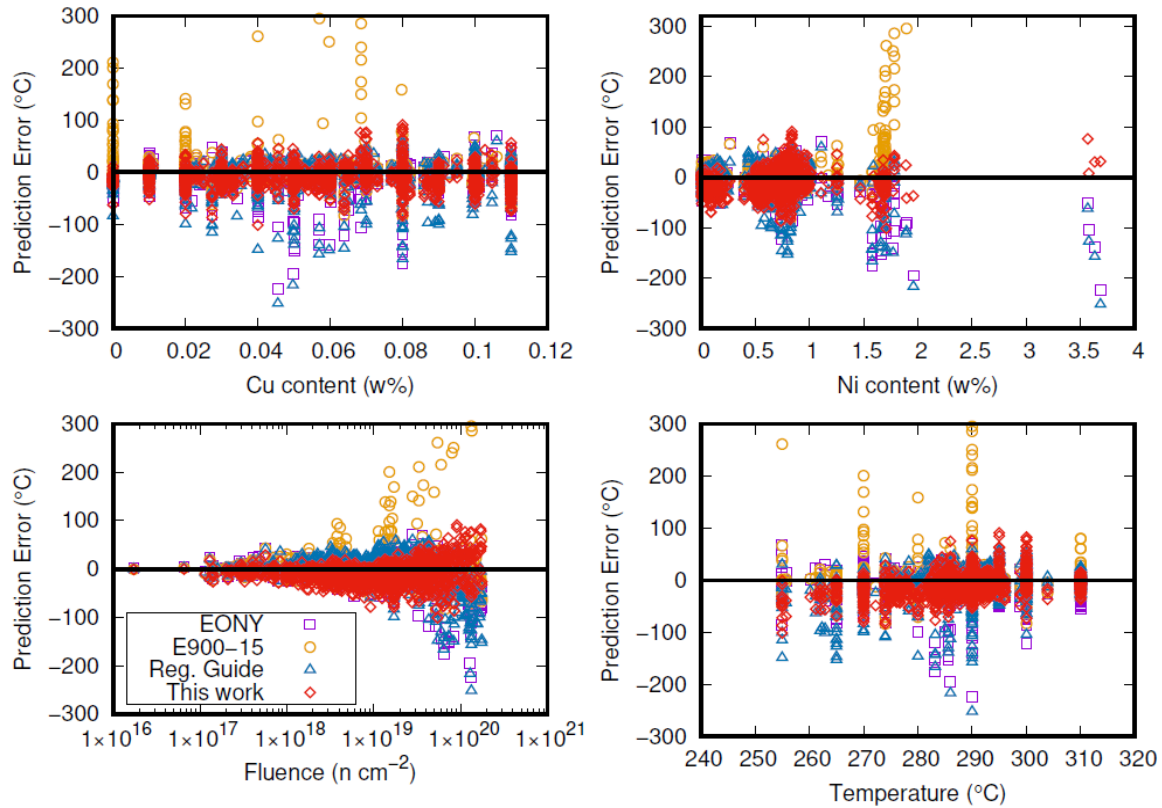

*Figure 9– Accuracy of prediction of ΔT achieved by the EONY embrittlement correlation method [62], The Regulatory Guide 1.99 Rev. 2 [63], and the ASTM E900-15 correlation [64], compared with the predictions achieved in this work. The panels plot the evolution of the prediction error with four key variables of the model.*

The success of our model strongly suggests that the effect of solute clusters dissolved in the bulk of grains is indeed the dominant factor to cause radiation-induced hardening and embrittlement in RPV steels. It also suggests that the fundamental physical mechanisms introduced in the microstructure evolution model are correct, or at least sufficient, to provide fair predictions of the density, size and volume fraction of these clusters. It is however important to discuss the limits of the assumptions made:

a) The ensuing model is here applied up to 0.1 at% Cu. As discussed above, higher Cu contents imply an increasingly dominant effect of precipitation as dictated by thermodynamics. Consequently, the embrittlement evolution with dose should rapidly saturate, a trend that the model in its present form cannot predict.

b) Irradiation doses well beyond those of relevance to RPV steel operation are known to induce the development of other microstructural features that influence their mechanical properties. For example, near or above 1 dpa, ferritic-martensitic steels develop dislocation loops and nanovoids, which also act as obstacles to the motion of dislocations [65]. While the dispersed barrier hardening theory still holds, these additional obstacles are not explicitly accounted for in the present model: embrittlement laws that superpose different defect populations should be applied.

c) If the P concentration is high enough, its segregation at grain boundaries is known to favour intergranular fracture [66], leading to embrittlement without hardening. This cannot be described by the dispersed barrier hardening model adopted here. Such effects should be assessed separately from the effect of bulk precipitation.

We should emphasize that a similar microstructure evolution model as described here was previously applied with success to predict the evolution of solute cluster volume fraction in model alloys for F/M steels [54], as a function of the received neutron dose. This results further confirms the robustness of the assumptions made here in terms of microstructure evolution mechanisms, which extend from RPV to F/M steels. That is, the mechanism of solute cluster formation can be considered essentially identical in RPV, VVER, and F/M materials and is driven by single point-defect that drag solutes towards sinks. The sinks that dominantly provide nucleation sites for solute clusters are irradiation-generated point-defect clusters. Thus, the radiation-induced character of the process is dominant over the radiation-enhanced one. A quantitative prediction of $\Delta T$ in F/M steels based on this sole information like for RPV steels is most likely not possible, for at least two reasons: (1) the significant difference in composition and microstructure between RPV and F/M steels, and between different types of F/M steels, which produces features such as secondary phases and makes it difficult to define the actual grain sizes that need to be introduced in the underlying hardening law (see, e.g., Ref. [59]); (2) unlike the majority of RPV and VVER steels, nanovoids and dislocation loops are not negligible in the case of F/M materials, especially considering that (much) higher neutron doses are generally reached there (> 1dpa in contrast to < 0.3 dpa for RPV and VVER materials). Discussing these aspects in detail goes beyond the scope of this paper. Nevertheless, our OKMC model is clearly applicable to predict the kinetics of solute clustering in any ferritic material relevant for nuclear application, i.e., RPV, VVER, and F/M steels, regardless of their intrinsic differences. This is an important point in support of the generality of the fundamental physical mechanisms that drive the microstructure evolution under irradiation.

Another important consideration stemming from our model is that the direct use of microstructural information in embrittlement correlations, and damage models in general, seems to significantly simplify their mathematical form: this can be appreciated by comparing Eq (3) and Eq (4) with the mathematical expression of, e.g., the EONY model or other correlations [42]. The whole dependence of radiation embrittlement on chemical composition, temperature, dose and dose-rate is fully accounted for by $N$ and $D$ and $f$, which are computed applying a microstructure evolution theory. The rest of the mathematical expression is just a coefficient. The present work thus suggests that, even though the long-term goal of multiscale approaches [7] is to predict materials behavior with models that describe all fundamental processes along increasing time and space scales, it is also possible to improve the reliability of engineering correlations by directly coupling them with microstructure evolution theories. Specifically, in this work we produced a tool that provides reasonable estimates of microstructural variables, starting from information that is always available (steel composition, temperature, dose and dose-rate), without the need to perform long and costly microstructural examinations (e.g., APT), and enabling the direct assessment of mechanical properties. We argue that, with the help of modern techniques, such as machine learning, this approach can be directly used for engineering purposes, by training an artificial intelligence model to provide similar, but faster, predictions of $N$ and $D$ or $f$, given the steel and the irradiation conditions [67]. Alternatively, this microstructural information can be used to feed more refined correlations with embrittlement, perhaps better rooted in physics, leading to even further reduced uncertainty in the prediction of $\Delta T$.

# 5 Conclusion

We have developed a robust microstructure evolution theory and model that quantitatively predicts the formation of solute clusters under irradiation in ferritic steels given their composition and the irradiation conditions. This piece of microstructural information proves to be sufficient to assess in a very simple way embrittlement as a function of the nuclear power plant operation time. The embrittlement of ferritic steels under irradiation is therefore dominated by the formation of nanometric solute clusters. The simple embrittlement law provided here in Eq. (3) and Eq (4), coupled with our microstructure evolution model for the assessment of solute clustering, proves to be even more reliable than existing correlations that are currently used to assess the integrity of reactor pressure vessel steels beyond the planned surveillance programs.

# Acknowledgements

This work received partial financial support in the framework of the 2019-2020 Euratom research and training programme (ENTENTE project, grant No 900018), as well as from the 2014-2018 Euratom research and training programme (SOTERIA project, grant No. 661913, and M4F project, grant No. 755039). It is also partly indebted to earlier 7$^{th}$ framework programme projects (LONGLIFE project, grant No. 249360, and MATISSE project, grant No. 604862). Finally, this work contributes to the Joint Programme on Nuclear Materials of the European Energy Research Alliance (EERA JPNM). The views and opinions expressed herein do not necessarily reflect those of the European Commission.

N. Castin acknowledges Prof. M. Miralles from the Buenos Aires Catholic university (UCA) for a critical review of this work.

# Author contributions

N.C., G.B. and L.Ma. led the research. L.Ma. developed the fundamental concepts of the theory. N.C. developed the object kinetic Monte Carlo (OKMC) model. L.Me., C.D. and A.B. derived the ab initio data for the development and parametrization of the OKMC model. G.B. and M.I.P. performed atomistic modelling studies that were used to parameterize the OKMC model described in this work. M.K. performed positron annihilation spectroscopy experiments that contributed to the OKMC model parametrization. C.C. performed a literature survey of atom probe tomography and embrittlement data, cited and used in this work. B.R. and B.G.F. performed atom probe tomography experiments. M.K., F.B. and J.M.H. significantly contributed to the development of the embrittlement model proposed in this work. M.S. supervised the gathering of the experimental data of radiation embrittlement in steels which were used to validate the embrittlement model proposed in this work. All authors contributed to the writing of the manuscript.